\documentclass[10pt,  conference]{IEEEtran}
\usepackage{todonotes}
\usepackage[nocompress]{cite}
\usepackage{url}
\usepackage{enumitem}
\usepackage{epstopdf}
\usepackage{graphicx}
\usepackage{caption}
\usepackage{amsmath}
\usepackage{tabu}
\usepackage{multirow}
\usepackage{adjustbox}
\usepackage{fontawesome}
\usepackage{algpseudocode}
\usepackage{soul}
\usepackage[bookmarks=false]{hyperref}
\usepackage{footnote}
\usepackage{lipsum}
\usepackage[ruled,vlined, linesnumbered]{algorithm2e}
\usepackage{subcaption}
\usepackage{listings}
\usepackage{booktabs}
\usepackage{tikz}
\usetikzlibrary{positioning}
\soulregister\cite7
\soulregister\ref7
\soulregister\pageref7
\soulregister\secref7
\soulregister\tabref7
\newcommand*\circled[1]{\tikz[baseline=(char.base)]{
            \node[shape=circle,draw,inner sep=0.75pt, text=white,fill=black] (char) {#1};}}
\newcommand\copyrighttext{%
  \footnotesize \copyright 2022 IEEE. Personal use of this material is permitted. Permission from IEEE must be obtained for all other uses, in any current or future media, including
reprinting/republishing this material for advertising or promotional purposes, creating new
collective works, for resale or redistribution to servers or lists, or reuse of any copyrighted
component of this work in other works.}
\newcommand\copyrightnotice{%
\begin{tikzpicture}[remember picture,overlay]
\node[anchor=south,yshift=10pt] at (current page.south) {\fbox{\parbox{\dimexpr\textwidth-\fboxsep-\fboxrule\relax}{\copyrighttext}}};
\end{tikzpicture}%
}

\definecolor{ao}{rgb}{0.0, 0.5, 0.0}
\definecolor{db}{rgb}{0.2, 0.2, 0.6}
\definecolor{cadmiumgreen}{rgb}{0.0, 0.42, 0.24}

\newcommand{\tabref}[1]{Table~\ref{#1}}

\newcommand{\secref}[1]{Section~\ref{#1}}

\SetAlFnt{\footnotesize}

\begin{document}
\title{Bunk8s: Enabling Easy Integration Testing of Microservices in Kubernetes}

\author{\IEEEauthorblockN{Christoph Reile\IEEEauthorrefmark{1}, Mohak Chadha\IEEEauthorrefmark{1}, Valentin Hauner\IEEEauthorrefmark{2}, Anshul Jindal\IEEEauthorrefmark{1}, Benjamin Hofmann\IEEEauthorrefmark{2}, Michael Gerndt\IEEEauthorrefmark{1}\\}
\IEEEauthorblockA{\IEEEauthorrefmark{1}Chair of Computer Architecture and Parallel Systems, Technische Universit{\"a}t M{\"u}nchen \\
Garching (near Munich), Germany \\ \IEEEauthorrefmark{2}MaibornWolff GmbH, Germany \\}
Email: \IEEEauthorrefmark{1}\{firstname.lastname\}@tum.de, \IEEEauthorrefmark{2}\{firstname.lastname\}@maibornwolff.de}

\maketitle
%Uncomment before final submission
% \pagenumbering{gobble}
\copyrightnotice
%  \copyrightnotice


% \input{sections/00_abstract}
\begin{abstract}

Microservice architecture is the common choice for cloud applications these days since each individual microservice can be independently modified, replaced, and scaled. However, the complexity of microservice applications requires automated testing with a focus on the interactions between the services. While this is achievable with end-to-end tests, they are error-prone, brittle, expensive to write, time-consuming to run, and require the entire application to be deployed. Integration tests are an alternative to end-to-end tests since they have a smaller test scope and require the deployment of a significantly fewer number of services. The de-facto standard for deploying microservice applications in the cloud is containers with Kubernetes being the most widely used container orchestration platform. To support the integration testing of microservices in Kubernetes, several tools such as Octopus, Istio, and Jenkins exist. However, each of these tools either lack crucial functionality or lead to a substantial increase in the complexity and growth of the tool landscape when introduced into a project. To this end, we present \emph{Bunk8s}, a tool for integration testing of microservice applications in Kubernetes that overcomes the limitations of these existing tools. \emph{Bunk8s} is independent of the test framework used for writing integration tests, independent of the used CI/CD infrastructure, and supports test result publishing. A video demonstrating the functioning of our tool is available from \url{https://www.youtube.com/watch?v=e8wbS25O4Bo}.

\end{abstract}


% \IEEEpeerreviewmaketitle
% \copyrightnotice
% \thispagestyle{empty}

% \todochris{Write details about the supervisors from MaibornWolff Valentin, Benjamin in the author section}
% Valentin Hauner, valentin.hauner@maibornwolff.de, Senior Software Engineer at MaibornWolff GmbH since 2016, Munich
% Benjaming Hofmann, benjamin.hofmann@maibornwolff.de, Senior Lead Test Architect at MaibornWolff GmbH since 2018, Munich

\section{Introduction}
\label{sec:intro}

Microservices-based applications are typically composed of independently deployable components that can be directly translated into services and automatically scaled on-demand. Components are independently replaceable and upgradable units of software~\cite{fowlermic}, consisting of several modules~\cite{clemson}. Linux containers, due to their portability, isolation, and high availability, have become the de-facto standard for developing, testing, and deploying such applications in cloud environments~\cite{cloud_container}. This is because containers enable users to package their application and its custom software dependencies as a single unit into easy-to-deploy images. Moreover, containers are a natural fit for microservices-based applications in the cloud due to their smooth integration with container orchestration platforms~\cite{kubesphere} for efficient resource management.  Recent trends~\cite{datadogreport} show that nearly $90$\% of all deployed containers are orchestrated, with Kubernetes (k8s)~\cite{kuber} being the most utilized container orchestration platform~\cite{sysdig}.

% Applications based on a microservice architecture are structured as a set of loosely coupled, independently deployable, and collaborating services~\cite{microservicearch} that can be automatically scaled on-demand. 

Developers can deploy k8s on-premise or can utilize the services offered by most commercial cloud providers such as Microsoft (Azure Kubernetes Service)~\cite{aks}. If a microservice-based application is deployed to a k8s cluster, it may rely on backend services that run outside the k8s cluster. For instance, services that provide storage or networking functionalities. For commercial cloud providers, these services are self-managed, with only the service's live API available for the developer to access. As a result, testing the interactions of modules from a microservice application with the modules from the backend service requires the entire backend service to be deployed. It cannot be done in isolation with only specific modules of the backend service. Therefore, testing the interactions of a microservice application with backend services can only be accomplished by running end-to-end tests or broad integration tests. Note that, integration tests can be classified into narrow and broad depending on their granularity (\S\ref{sec:integeration_testing}). 

End-to-end tests are brittle, expensive to write, time-consuming to execute, and require the deployment of the entire application~\cite{fowlerpyr}. Broad integration testing of microservices in k8s can be done in two ways, i.e., from inside or outside the cluster. Running broad integration tests for microservices from outside the k8s cluster has several drawbacks, making them as inefficient as end-to-end tests. For example, the lack of public interfaces in the services which allow them to be accessed from outside the k8s cluster. As a result, the standard way of broad integration testing of microservices is from inside the k8s cluster. For the integration testing of microservices from inside the k8s cluster, several tools such as Octopus~\cite{octopus}, Istio~\cite{istio}, and Jenkins~\cite{jenkins} exist. However, the existing tools lack crucial functionalities or are dependent on particular infrastructure for the execution of these tests. To this end, we present \emph{Bunk8s} that overcomes the limitations of these existing tools.

% For this, several tools such as Octopus, Istio, and Jenkins exist. However, 

% \todomc{Work on intro later}

Towards easy integration testing in Kubernetes, our key contributions are:
\begin{itemize}
    \item We implement and present \emph{Bunk8s}\footnote{https://github.com/Teiktos/bunk8s}, a tool for broad integration testing of microservices in Kubernetes.
    \item \emph{Bunk8s} is independent of the test framework used for writing integration tests, independent of the used CI/CD infrastructure, and supports publishing of test results. Thus, making the inclusion of \emph{Bunk8s} in a project relatively easy.
    \item We demonstrate the functioning of \emph{Bunk8s} by running broad integration tests for a real-world microservice application on a commercial cloud provider, i.e., Azure.
    
\end{itemize}

The rest of the paper is structured as follows. \S\ref{sec:integeration_testing} provides a background on integration testing. In \S\ref{sec:bunks}, the system design of \emph{Bunk8s} and its workflow for integration testing of microservices are described. In \S\ref{sec:case_study}, we demonstrate the functioning of \emph{Bunk8s} for the integration testing of a real-world microservice application. \S\ref{sec:whybunk8s} motivates the need for \emph{Bunk8s} and presents a comparison with other tools for integration testing of microservices. Finally, \S\ref{sec:futurework} concludes the paper and presents an outlook.
\section{Background}
\label{sec:integeration_testing}

\begin{figure}[t]
    \centering
    \includegraphics[width=0.55\columnwidth]{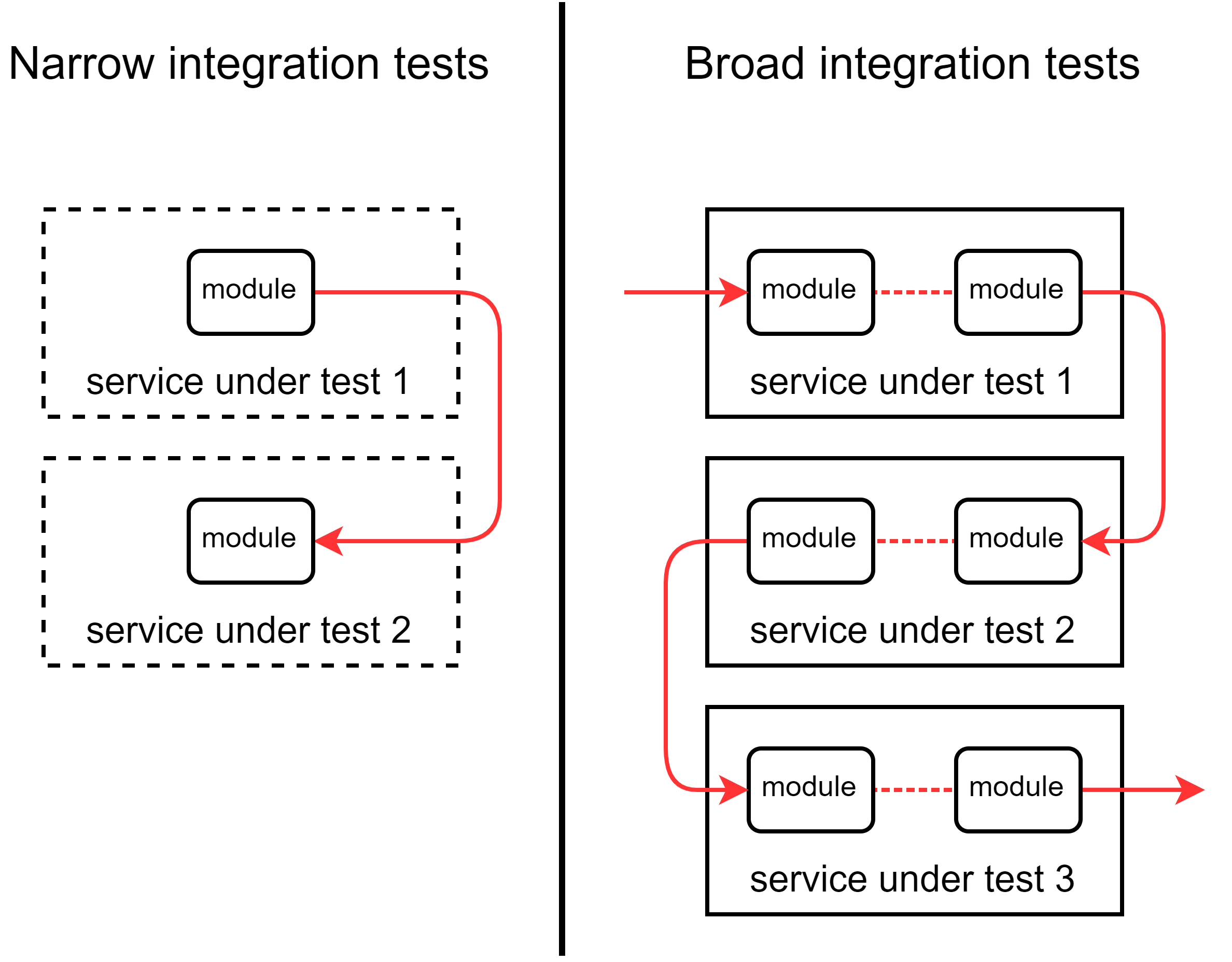}
    \caption[Room Tracking System Components]{Narrow and Broad integration testing.}
    \label{fig:narrow_broad_int_tests}
\end{figure}

% In microservice architectures, software systems are broken down into independently deployable components, which can directly be translated into microservices. They are independently replaceable and upgradable units of software \cite{fowlermic}, consisting of several modules \cite{clemson}. 

% Integration tests are "typically used to verify interactions between layers of integration code and the external microservices to which they are integrating"\cite{clemson}. They can be further categorized by their granularity, into narrow and broad integration tests. The former refers to only testing the modules of microservice, that communicate with each other. 
Integration tests are generally used to verify the interactions between the layers of integration code and the external microservices to which they are integrating~\cite{clemson}. They can be further categorized by their granularity, into narrow and broad integration tests. Narrow integration tests refer to testing only the modules of microservices that communicate with each other. Therefore, they only require instantiating 
the modules under test~\cite{fowlerinttest}. Microservice applications that run in commercial cloud offerings of k8s may integrate with services that run outside of k8s, e.g.,  Azure Cosmos DB, Azure Service Bus. Such services only provide developers with their live versions and it is not possible to instantiate only modules of these services that can integrate with the components of the microservice application. As a result, testing the interaction of microservices and services external to Kubernetes is not possible with narrow integration tests. On the other hand, broad integration tests refer to testing code paths through several services and all of their modules. Moreover, broad integration tests can be of varying granularity, depending on the scope of the subsystem under test~\cite{clemson}. They require live versions of the components under test, a substantial test environment and network access \cite{fowlerinttest}. The difference in the code paths used for writing narrow and broad integration tests is shown in Figure~\ref{fig:narrow_broad_int_tests}.

\section{Bunk8s}
\label{sec:bunks}

% In this section, the functional and non-functional requirements on Bunk8s will be elicited and the system design will be presented using a component diagram. Afterwards, the technical implementation of Bunk8s will be explained and elaborated on by describing step-by-step how a testing process with Bunk8s works.
In this section, we first describe our design goals for the development of \emph{Bunk8s}. Following this, we present its system design. Finally, we describe in detail, the workflow of \emph{Bunk8s} for the broad integration testing of microservices in k8s.

\subsection{Design Goals}

To facilitate the ease-of-use for utilizing \emph{Bunk8s} in a project and to overcome the limitations of the current tools for the integration testing of microservices in k8s (\S\ref{sec:whybunk8s}), we chose the following design goals:
\begin{itemize}
    \item Independence of the testing framework used for writing and generating tests.
    \item Independent of the specific CI/CD tools used within a project.
    \item Compatible with different container runtimes such as containerd~\cite{containerd} and CRI-O~\cite{crio}.
    \item Portable across the different Kubernetes offerings by the commercial cloud providers such as AKS~\cite{aks}, GKE~\cite{gke}.
    \item Support for publishing of test results.
\end{itemize}

\subsection{System Design}
\label{sec:sysdesign}
\begin{figure}[t]
        \centering
        \includegraphics[width=0.65\columnwidth]{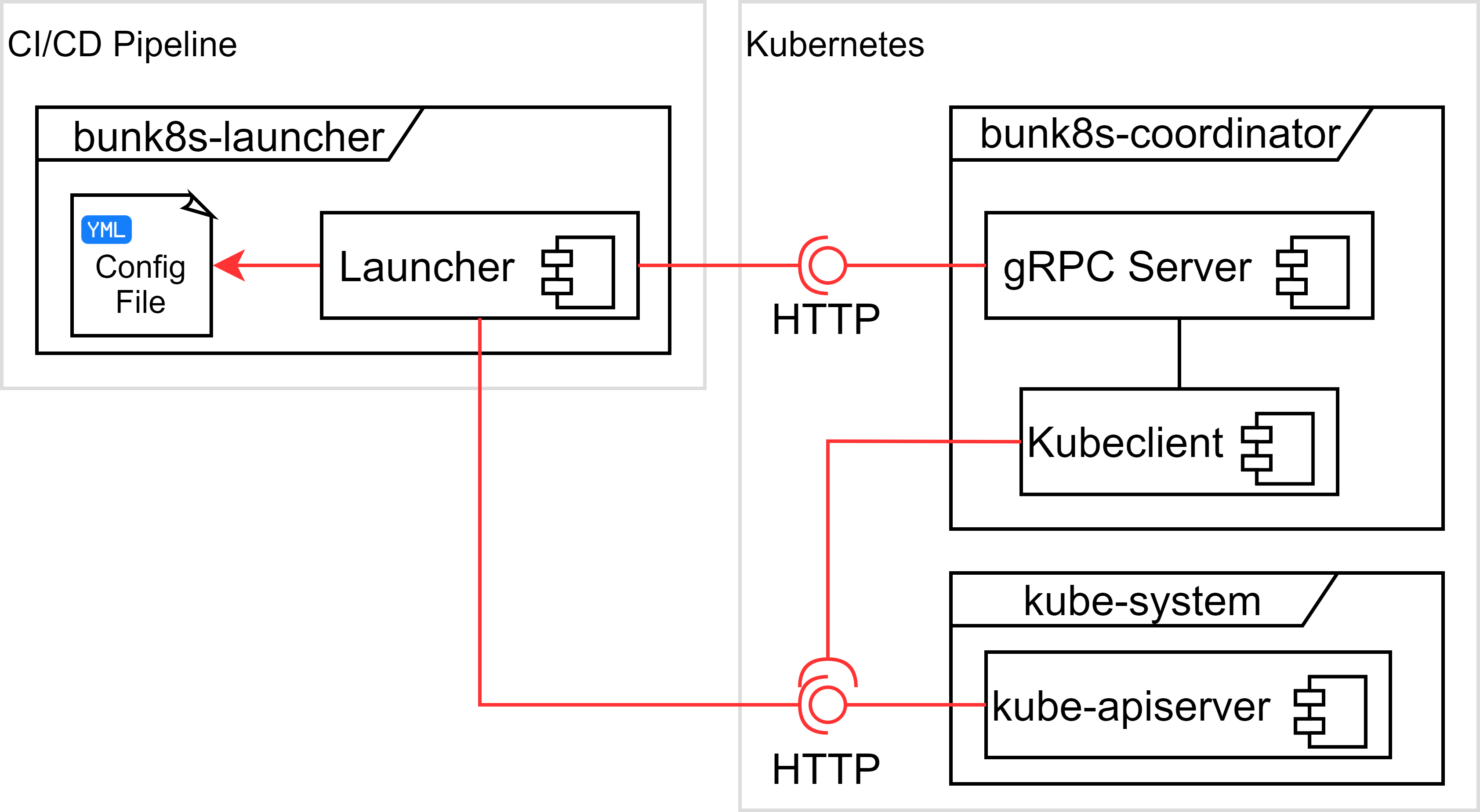}
        \caption[Bunk8s components]{Component diagram of \emph{Bunk8s}.}
        \label{fig:bunk8s_component}
    \end{figure}

% The underlying approach with Bunk8s is to create a tool, that allows deploying test runner containers in test runner pods to a testing namespace. In this namespace, the services under test are deployed. The test runner containers will then execute the tests. The setup of cloud infrastructure and the deployment of the services under test is not a functionality of Bunk8s and should be handled by tools that are suited for that task, e.g. Helm or Terraform\footnote{https://www.terraform.io/}.
\emph{Bunk8s} allows deploying test runner containers in test runner pods to a testing namespace. In this namespace, the microservices under test are deployed. Note that, test runner containers are written by developers and contain the integration tests for the microservice application. Following this, the tests are executed by the test runner containers. Setting up of the cloud infrastructure and the deployment of the microservices under test is not a functionality of \emph{Bunk8s} and should be done with tools such as Helm~\cite{helm} or Terraform~\cite{terraform}. The different components of \emph{Bunk8s} are shown in Figure~\ref{fig:bunk8s_component}. It consists of two components, i.e., \circled{1} the Launcher and \circled{2} the Coordinator. The Launcher runs in the CI/CD pipeline while the Coordinator runs in the k8s cluster.
% It consists of three components, i.e., \circled{1} the Launcher, \circled{2} the Coordinator, and \circled{3} the Test Runner. All components of \emph{Bunk8s} are implemented using \texttt{Go}.

    %   \begin{figure}[t]
    %     \centering
    %     \includegraphics[width=\columnwidth]{figures/example_application_testrun_communication.png}
    %     \caption[Bunk8s components]{Component diagram of Bunk8s}
    %     \label{fig:bunk8s_component}
    % \end{figure}
% Bunk8s will consist of two components, a launcher running in the pipeline and a coordinator running in the Kubernetes cluster. A component diagram is given in Figure \ref{fig:bunk8s_component}.

\subsubsection{The Launcher}
\label{sec:launcher}

% The Launcher runs outside of Kubernetes as part of the continuous integration process in a pipeline. 
% After the container images of the services under test are built and pushed to a container registry by the pipeline, the Bunk8s launcher is started, which starts the testing process.
% It loads a configuration file, which provides the information that is required by the launcher in order to connect to the coordinator. This includes the IP or URL of the Kubernetes ingress that has the Bunk8s coordinator configured as the backend service. It also provides information required by the controller for the test run. This includes names of the test runner images, in which namespace to deploy the test runner pods, the name of the test runner pods and a timeout.
% The Launcher connects to the coordinator via HTTP and sends the data.
% After the testing is finished and the coordinator responds to the launcher, the launcher calls the \textit{kube-apiserver} and extracts the test results from the test runner containers via HTTP and stores them on the pipeline runner. After the extraction, the test results can be published with the pipelines default test result publishing functionality, e.g. GitLab artifacts.
% The launcher will be designed to run in a container, e.g. as a Docker container or another arbitrary container runtime, making it platform independent and easily adaptable to other container runtimes.

The Launcher runs outside of k8s as part of the continuous integration (CI) process in a pipeline. After the container images of the microservices under test are built and pushed to a container registry by the pipeline, the \emph{Bunk8s} launcher is started, which in turn starts the testing process. The launcher takes a \texttt{YAML} configuration file as input, an example of which is shown in Listing~\ref{lst:config_file}. The configuration file provides the relevant information that is required by the launcher for connecting to the coordinator (\S\ref{sec:coordinator}), i.e., its IP address and port number. The developers can also provide a TLS certificate file for an authorized connection that is placed in the container image of the launcher. The configuration file also includes information required by the coordinator for running the integration tests. This includes the names of the test runner images, the namespace to which the test runner pods should be deployed, the name of the test runner pods, and a test timeout value. Note that, the list \texttt{testRunnerPods} allows defining multiple test runner pods which can be deployed in different namespaces, while the list \texttt{containers} allows defining multiple test runner containers per pod. The launcher connects to the coordinator via HTTP and sends the configuration data. After the testing has finished, the coordinator responds to the launcher. Following this, the launcher calls the \textit{kube-apiserver} and extracts the test results from the test runner containers via HTTP, and stores them on the pipeline runner. Pipeline runner is an agent that executes the CI/CD workflow. After the extraction, the test results can be published with the pipeline's default test result publishing functionality such as GitLab artifacts~\cite{gitlabar}. The launcher is designed to run in a container which makes it platform-independent and easily adaptable to other container runtimes.

\lstset{ %
    basicstyle=\ttfamily\footnotesize,
    commentstyle=\color{blue}\ttfamily,
    frame=single,
    language=Bash,
    showstringspaces=false,
    numbers=left,
    xleftmargin=2.6em,
    framexleftmargin=2.7em,
    morekeywords={blue},
    morestring=[s][\color{Gray}]{<}{>}
}

\begin{lstlisting}[caption={
   Example configuration file taken as input by the \emph{Bunk8s} launcher.
}, float, floatplacement=t, captionpos=b, basicstyle=\ttfamily\tiny,  belowskip=-1.5 \baselineskip, frame=single,  numbers=left,breaklines,language=Bash,    xleftmargin=2.6em,
    framexleftmargin=2.7em, label={lst:config_file}]
launcherConfig:                 
    coordinatorIp:          # IP or URL of the Ingress
    coordinatorPort:        # Port on the Ingress
    certFile:               # Name of the CAs .pem certificate file
coordinatorConfig:
    testRunnerPods:   
        - podName:          # Name of a test runner pod
          namespace:        # Namespace the test runner pod is deployed in
          testTimeout:      # Timeout for the test runner pod
    containers:
        - containerName:    # Name of a test runner container
          image:            # Name of a test runner container image
          startupCommands:  # Startup commands for the test runner container image. Replaces the images entrypoint.
          startupCommandsArgs: # Arguments for the startup commands. 
          testResultPath:   # Directory inside the test runner container image, where the test results are stored.
\end{lstlisting}

% After the testing is finished and the coordinator responds to the launcher, the launcher calls the \textit{kube-apiserver} and extracts the test results from the test runner containers via HTTP and stores them on the pipeline runner. 

\subsubsection{The Coordinator}
\label{sec:coordinator}

The \emph{Bunk8s} Coordinator handles the deployment of the test runner containers and watches their state until they have finished. It consists of two sub-components. First, an RPC server for communication and synchronization between the launcher and the coordinator. Second, a k8s clientset that connects to the \textit{kube-apiserver} via HTTP, sends a request to create the test runner pods, and watches their state. The coordinator's RPC server is exposed to the launcher via an Ingress. On test completion, the coordinator sends a response to the Launcher, containing the information that is required to extract the test results from the test runner containers. The test runner container image that is deployed in a pod by the \emph{Bunk8s} coordinator must be built and pushed to a container registry from which k8s can pull container images. Since the integration tests are executed inside a container that is separate from the other components of \emph{Bunk8s}, the developers can use any test framework or programming language to write their tests.

% An RPC server, which allows the synchronization of the test extraction by the launcher, and a Kubernetes clientset, that connects to the \textit{kube-apiserver} via HTTP and sends a request to create the test runner pods and watches the test runner pod's state. The coordinator's RPC server is exposed to the launcher via an Ingress.
% After the test runner pods finish, the coordinator sends a response to the Launcher, containing the data that is required to extract the test results from the test runner containers. 

% \subsubsection{The Test Runner}

% % The test runner container image that is deployed in a pod by the coordinator must be built and pushed to a container registry, from whichKubernetes can pull container images. Since the tests are run in a container that is separated from the other components of Bunk8s, it isnot required to use a specific test framework or programming language to write tests. 

% The test runner container image that is deployed in a pod by the \emph{Bunk8s} coordinator must be built and pushed to a container registry from which k8s can pull container images. Since the integration tests are executed inside a container that is separate from the other components of \emph{Bunk8s}, the developers can use any test framework or programming language to write their tests.

\begin{figure}[t]
    \centering
    \includegraphics[width=\columnwidth]{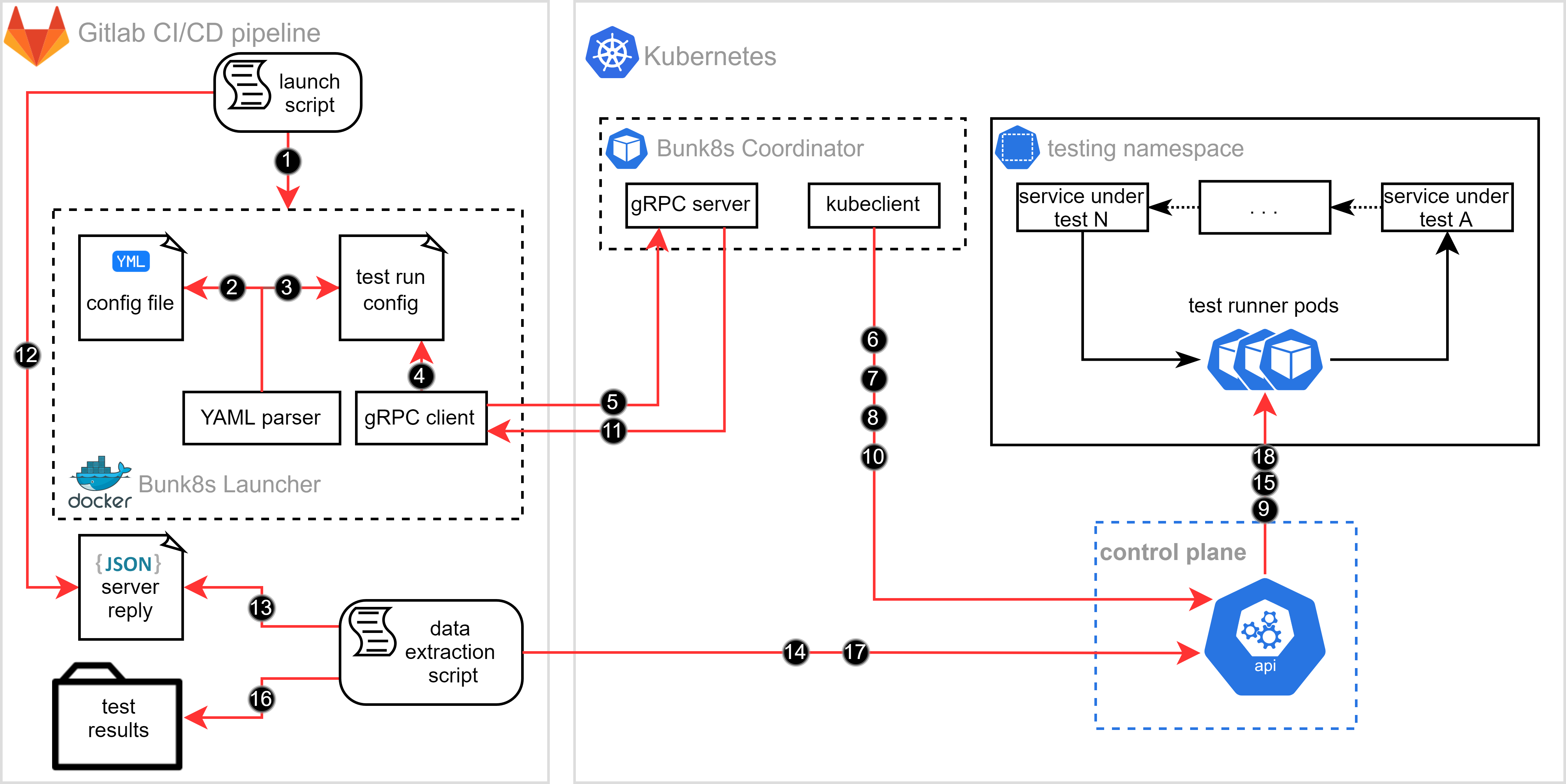}
    \caption[Bunk8s components]{Workflow of Integration testing with \emph{Bunk8s}.}
    \label{fig:bunk8s_workflow}
\end{figure}

\subsection{Integration testing with Bunk8s}
\label{sec:testingworkflow}
To install \emph{Bunk8s} in a k8s cluster, the launcher and coordinator (\S\ref{sec:sysdesign}) images must be built using \texttt{Dockerfiles}. The launcher image must be uploaded to a container registry from where it can be pulled by the pipeline runners, while the coordinator image must be uploaded to a registry to which the k8s cluster has access. For a quick setup of the \emph{Bunk8s} coordinator, we provide a pre-configured Helm chart in our repository (\S\ref{sec:intro}). Note that, the coordinator can be deployed in any k8s namespace without affecting its functionality. However, it is recommended that the coordinator is deployed in a namespace of its own. The project admin is responsible for creating the role-based access control rules (RBAC) for the coordinator pod in its namespace. The detailed workflow of the testing process with \emph{Bunk8s} is shown in Figure~\ref{fig:bunk8s_workflow}. 

At the beginning \circled{1}, the \emph{Bunk8s} launcher container is started by using a launch script. The launcher takes a \texttt{YAML} configuration file as input (\S\ref{sec:sysdesign}) which is mounted into the container at startup. The \texttt{YAML} parser parses the configuration file, checks if it is valid \circled{2}, and creates a \texttt{Go} struct from the configuration data \circled{3}. Following this, the \texttt{gRPC} client is created that receives the test run configuration created in the previous step \circled{4}. The \texttt{gRPC} client calls the \texttt{DeployTestRunner} RPC method registered in the \texttt{gRPC} server that invokes several functions defined in the kubeclient package \circled{5}. For brevity, those function names have been omitted from Figure~\ref{fig:bunk8s_workflow}. As part of the RPC method in \circled{5}, the kubeclient does several HTTP requests to the kube-apiserver. First, it does a \texttt{GET} request for each namespace provided in the configuration data \circled{6}. After it receives a response, it checks if the given namespaces already exist or not. If not, the test run is terminated and the coordinator sends an error code as a \texttt{gRPC} reply to the launcher. Second, it does a \texttt{GET} request for each pod from the configuration data \circled{7}. On receipt of the response, it checks whether the pods already exist within the namespace that they should be scheduled in. If the pods do not exist, the kubeclient creates k8s API pod objects. Third, it does a \texttt{POST} request for each pod object created in the previous step \circled{8}. After this, the test runner pods are created by the kube-apiserver \circled{9}. Finally, the kubeclient creates a watcher for each test runner pod. After all the tests have finished, the \texttt{DeployTestRunner} RPC method ends \circled{10}. Following this, the \texttt{gRPC} server sends the server response to the \texttt{gRPC} client which is written to the logs of the \emph{Bunk8s} launcher container \circled{11}. The server reply is stored as a JSON file by the launch script (\circled{12}) and then read by the data extraction script (\circled{13}). After this, the data extraction script executes kubectl which does an HTTP \texttt{POST} request to the kube-apiserver to copy the test results from the test runner pods to the pipeline runner \circled{14}. The test results are extracted by the kube-apiserver from the sidecar containers inside the test runner pods \circled{15}. Following this, the test results are stored on the pipeline runner  \circled{16}. Finally, the data extraction script does an HTTP \texttt{DELETE} request to the kube-apiserver (\circled{17}) to delete the test runner pods (\circled{18}). The results of the integration tests can now be published using the CI pipeline's publishing mechanisms.

% test runner pods are deleted by the kube-apiserver 
%After all tests have finished

% \input{sections/04_evaluation}

\section{Case Study: Room Tracking System}
\label{sec:case_study}
% In this section, the successful implementation of Bunk8s will be demonstrated by
% running broad integration tests for a microservice example application. This serves as
% an example of how to use Bunk8s and shows that the requirements defined in chapter 4
% are met.
In this section, we demonstrate the usage of \emph{Bunk8s} for the broad integration testing of a real-world microservice application.

\subsection{Experimental Setup}
\label{sec:env}

% For this demonstration, a remote test environment is available. The environment includes GitLab and a Kubernetes namespace in an Azure Kubernetes Service (AKS), as well as an Azure Container Registry. GitLab is a DevOps platform, that provides remote Git repositories, as well as CI/CD pipelines \cite{ch5-gi-whatisgitlab}. The example application, as well as Bunk8s, are located in a GitLab Git repository and are deployed to the AKS with GitLab's built-in pipeline functionality.
To demonstrate the functioning of \emph{Bunk8s}, we create a Kubernetes cluster using the Azure Kubernetes Service (AKS)~\cite{aks}. We use GitLab as the DevOps platform since it provides remote Git repositories and CI/CD pipelines~\cite{ch5-gi-whatisgitlab}. As the registry for containers, we use the Azure Container Registry~\cite{acr}. The chosen application and \emph{Bunk8s} are deployed to the AKS by utilizing the pipeline functionality present in GitLab. Although we use AKS for our experiments, \emph{Bunk8s} is compatible with commercial k8s offerings of all cloud providers.

\subsection{Application Design}
\label{sec:design}

\begin{figure}[t]
    \centering
    \includegraphics[width=0.75\columnwidth]{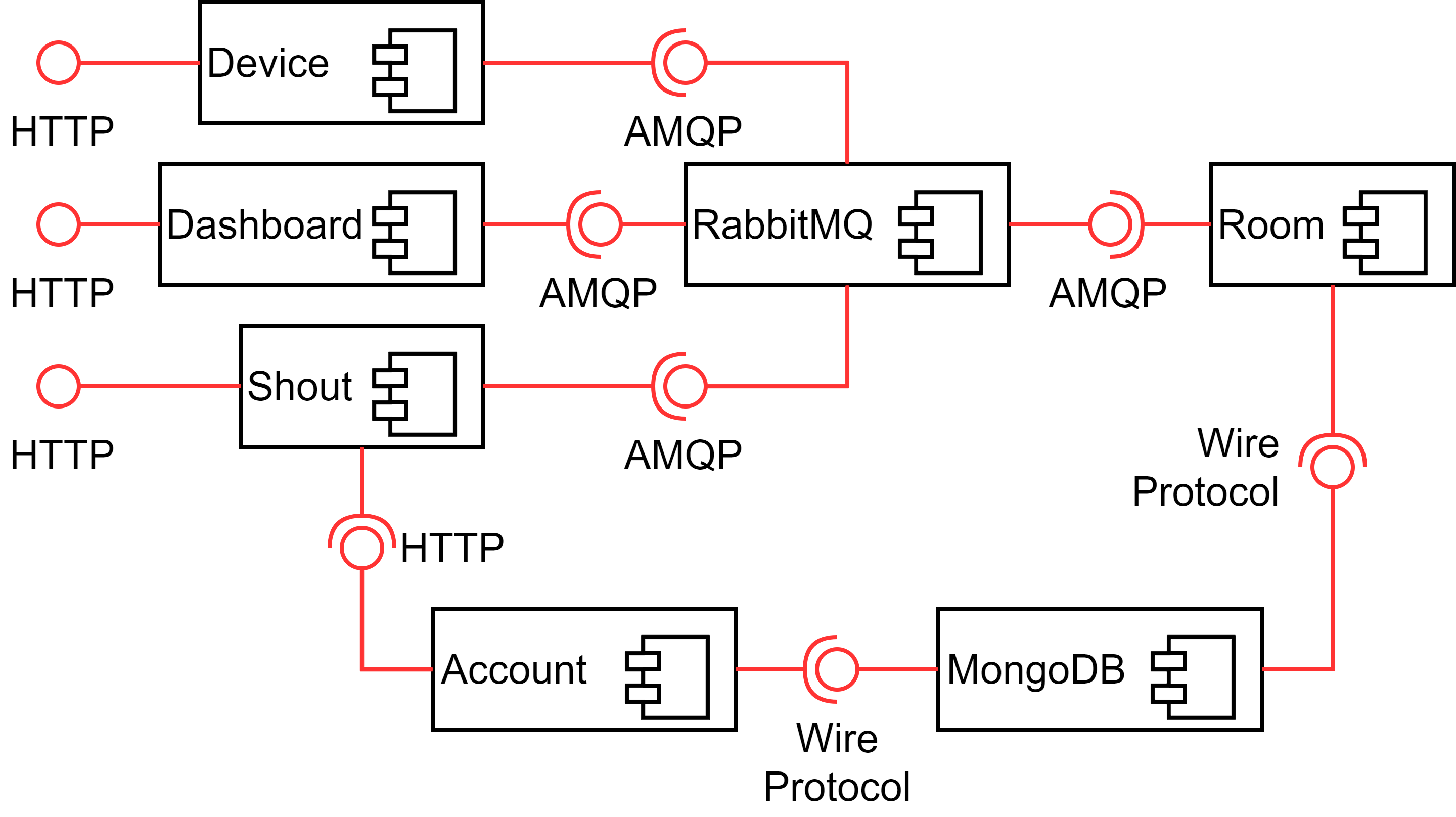}
    \caption[Room Tracking System Components]{Room Tracking Application Microservices.}
    \label{fig:example_components}
\end{figure}

% The example application is a room tracking system, that is used to train developers on the architecture of microservices. A user can hold a card up to a badge reader inside a room and that room will be marked as occupied by the application. Users can also access a dashboard via a browser, that gives an overview of the room utilization. Furthermore, when a user enters a room, it is announced on a shout website, separate from the dashboard. 
The application we use is a room tracking system. A user can hold a card up to a badge reader inside a room and that room will be marked as occupied by the application. Users can also access a dashboard via a browser that provides an overview of the room utilization. Furthermore, when a user enters a room, it is announced on a shout website that is separate from the dashboard.

Figure~\ref{fig:example_components} gives an overview of the different microservices present in the room tracking system. It consists of two frontend services, i.e., \textit{Dashboard} and \textit{Shout}, three backend services, i.e., \textit{Device}, \textit{Account}, and \textit{Room}, and two infrastructure services, i.e., \textit{RabbitMQ} and \textit{MongoDB}. The frontend services \textit{Dashboard} and \textit{Shout} provide the dashboard and shout websites. \textit{RabbitMQ} provides message queues for the communication between the \textit{Device}, \textit{Dashboard}, \textit{Shout} and the \textit{Room} service. When a user enters a room, a badge reader device sends a badge reader event to the \textit{Device} service. Following this, the \textit{Device} service creates a device event containing the card and room numbers. The device event is then sent to the \textit{Room} service via the message queue. The \textit{Room} service updates the room allocation in \textit{MongoDB} and creates a room event that describes which user entered or left the room and forwards it to the \textit{Dashboard} and \textit{Shout} services via the two different message queues. After receiving the room event, the \textit{Dashboard} service updates a counter that tracks the number of users that have checked into a room. On the other hand, the \textit{Shout} service requests the name of the user from the room event from the \textit{Account} service. The \textit{Account} service provides the functionality for creating, modifying, and listing the user accounts that are stored in the \textit{MongoDB} service. Note that, \texttt{AMQP}~\cite{standard2012oasis} is a message protocol for transferring messages, while the wire protocol is the default communication protocol used by MongoDB~\cite{wire}.

\subsection{Broad Integration Testing Results}
% Broad integration tests cover multiple code paths through several microservices and all  of their components (\S\ref{sec:integeration_testing}). As a result, multiple services of the room tracking application will be under test (\S\ref{sec:design}). Since \emph{Bunk8s} \hl{enables testing of backend subsystems without requiring the reconfiguration of ingresses in k8s, as well as the testing of backend services that do not provide an HTTP or HTTPS interface, the services \textit{Dashboard}, \textit{Shout}, and \textit{Account} are not tested and omitted from deployment}. As a result, the microservices under test are \textit{Device}, \textit{Room}, \textit{RabbitMQ}, and \textit{MongoDB}. 

Broad integration tests cover multiple code paths through several microservices and all  of their components (\S\ref{sec:integeration_testing}). As a result, multiple services of the room tracking application will be under test (\S\ref{sec:design}). Since \emph{Bunk8s} enables testing of backend services without requiring the reconfiguration of ingresses in k8s, as well as the testing of backend services that do not provide an HTTP or HTTPS interface, the services \textit{Dashboard}, \textit{Shout}, and \textit{Account} are not tested and omitted from deployment. To this end, the microservices under test are \textit{Device}, \textit{Room}, \textit{RabbitMQ}, and \textit{MongoDB}. 

% \begin{figure}[t]
% \centering
% \begin{minted}[frame=single, framesep=2mm, fontsize=\tiny,linenos,breaklines, obeytabs=true,tabsize=4, numbersep=-10pt]{text}

%     === RUN   TestDeviceServiceReceiveDeviceEvent
%     --- PASS: TestDeviceServiceReceiveDeviceEvent (0.27s)
%     === RUN   TestRoomServiceRoomEventResponse
%     === RUN   TestRoomServiceRoomEventResponse/Queue_Dashboard
%     === RUN   TestRoomServiceRoomEventResponse/Queue_Shout
%     --- PASS: TestRoomServiceRoomEventResponse (0.07s)
%         --- PASS: TestRoomServiceRoomEventResponse/Queue_Dashboard (0.00s)
%         --- PASS: TestRoomServiceRoomEventResponse/Queue_Shout (0.00s)
%     === RUN   TestMongoDBRoomAllocationEntry
%     --- PASS: TestMongoDBRoomAllocationEntry (0.09s)
%     PASS
%     ok  	showcase	0.435s    
    
% \end{minted}
% \caption{Test result logs after successful completion of the broad integration tests with \emph{Bunk8s}.}
% \label{lst:succresult}
% \end{figure}

\begin{lstlisting}[caption={
    Test result logs after successful completion of the broad integration tests with \emph{Bunk8s}.
}, float, floatplacement=t, captionpos=b, basicstyle=\ttfamily\tiny,  belowskip=-1.5 \baselineskip, frame=single,  numbers=left,     xleftmargin=2.6em,
    framexleftmargin=2.7em, label={lst:succresult}]
%  === RUN   TestDeviceServiceReceiveDeviceEvent
%  --- PASS: TestDeviceServiceReceiveDeviceEvent (0.27s)
%  === RUN   TestRoomServiceRoomEventResponse
%  === RUN   TestRoomServiceRoomEventResponse/Queue_Dashboard
%  === RUN   TestRoomServiceRoomEventResponse/Queue_Shout
%  --- PASS: TestRoomServiceRoomEventResponse (0.07s)
%  --- PASS: TestRoomServiceRoomEventResponse/Queue_Dashboard (0.00s)
%  --- PASS: TestRoomServiceRoomEventResponse/Queue_Shout (0.00s)
%  === RUN   TestMongoDBRoomAllocationEntry
%  --- PASS: TestMongoDBRoomAllocationEntry (0.09s)
%  PASS
%  ok  	showcase	0.435s   

\end{lstlisting}

% \begin{figure}[t]
% \centering

% \begin{minted}[frame=single, framesep=2mm, fontsize=\tiny,linenos,breaklines, obeytabs=true,tabsize=4, numbersep=-10pt]{text}

%     [...]
%     === RUN   TestMongoDBRoomAllocationEntry
%         main_test.go:152: Failed to create MongoDB client: Failed to ping MongoDB: server     selection error: context deadline exceeded, current topology: { Type: Unknown, Servers:     [{ Addr: mongodb://mongo.bunk8s-fe.svc.cluster.local, Type: Unknown, Last error: connection() error occured during     connection handshake: dial tcp mongodb://mongo.bunk8s-fe.svc.cluster.local: connect: connection refused }, ] }
%     --- FAIL: TestMongoDBRoomAllocationEntry (10.00s)
%     FAIL
%     ok  	showcase	10.101s

% \end{minted}
% \caption{Test result logs if a broad integration test failed with \emph{Bunk8s}.}
% \label{lst:notsuccresult}
% \end{figure}

\begin{lstlisting}[caption={
    Test result logs after successful completion of the broad integration tests with \emph{Bunk8s}.
}, float, floatplacement=t, captionpos=b, basicstyle=\ttfamily\tiny,  belowskip=-1.5 \baselineskip, frame=single,  numbers=left,breaklines,tabsize=4,     xleftmargin=2.6em,
    framexleftmargin=2.7em, label={lst:notsuccresult}]
[...]
=== RUN   TestMongoDBRoomAllocationEntry
     main_test.go:152: Failed to create MongoDB client: Failed to ping MongoDB: server     selection error: context deadline exceeded, current topology: { Type: Unknown, Servers:     [{ Addr: mongodb://mongo.bunk8s-fe.svc.cluster.local, Type: Unknown, Last error: connection() error occured during     connection handshake: dial tcp mongodb://mongo.bunk8s-fe.svc.cluster.local: connect: connection refused }, ] }
--- FAIL: TestMongoDBRoomAllocationEntry (10.00s)
FAIL
ok  	showcase	10.101s
\end{lstlisting}

%More explanation of the integration tests

For writing the integration tests, we use Go and its default testing package. We generate a badge reader event which is sent to the \textit{Device} service. Following this, the contents of the room events created by the \textit{Room} service are checked for validity. Finally, we check the validity of the room allocated in the \textit{MongoDB} database. As described in \S\ref{sec:testingworkflow}, the test results are extracted and stored in artifact store provided by GitLab. Listing~\ref{lst:succresult} shows the logs of the test results after all broad integration tests have completed successfully. \emph{Bunk8s} also extracts test results if a test fails. For instance, Listing~\ref{lst:notsuccresult} shows logs of the test \texttt{TestMongoDBRoomAllocationEntry} when no connection to the \textit{MongoDB} service could be established. As a result, the test result publishing supported by \emph{Bunk8s} allows developers to easily analyse and debug their applications.

% caption={Test results after successful Broad Integration tests with \emph{Bunk8s}},label={lst:succresult}, captionpos=b
% Since we use AKS (\S\ref{sec:env}) for our k8s cluster, we replace the \textit{MongoDB} and \textit{RabbitMQ} services by the Azure Cosmos DB and Azure Service Bus respectively. Both of these services run outside of the k8s cluster on the Azure cloud platform.
\label{sec:results}

\section{Why Bunk8s?}
\label{sec:whybunk8s}

\begin{table}[t]
\centering
\begin{adjustbox}{width=9cm,center}
\begin{tabu}{ |c|c|c|c|c| }
    \hline
    \textbf{Supported Features} & \textbf{Bunk8s} \textbf{[This Work]} & \textbf{Octopus}~\cite{octopus}  & \textbf{Istio}~\cite{istio}  & \textbf{Jenkins}~\cite{jenkins}  \\
    \tabucline{-}
    Broad integration testing & \faThumbsOUp & \faThumbsOUp & \faThumbsDown & \faThumbsDown\\
    \tabucline{-}
    Narrow integration testing & \faThumbsDown & \faThumbsDown & \faThumbsOUp & \faThumbsOUp\\
    \tabucline{-}
    Test framework independent & \faThumbsOUp & \faThumbsOUp & \faThumbsDown & \faThumbsOUp \\
    \tabucline{-}
    CI/CD Infrastructure independent & \faThumbsOUp & \faThumbsOUp & \faThumbsDown & \faThumbsDown\\
    \tabucline{-}
    Test result publishing & \faThumbsOUp & \faThumbsDown & \faThumbsOUp & \faThumbsOUp\\
    \tabucline{-}
\end{tabu}
\end{adjustbox}
\caption{Comparison of the different features supported by \emph{Bunk8s} with the other available tools for integration testing. \faThumbsOUp Supported. \faThumbsDown No support. }
\label{table:comparison}
\end{table}

To support the integration testing of microservices inside k8s several tools exist. These include Octopus~\cite{octopus}, Istio~\cite{istio}, and Jenkins~\cite{jenkins}. Octopus allows developers to define integration tests with custom resource definitions~\cite{crd}. For each test, Octopus deploys a pod to k8s. Furthermore, it supports re-running tests to prevent flaky test results. The Istio testing framework allows developers to test microservice applications that use the Istio service mesh to handle the communication between microservices. Developers can also use Jenkins to run integration tests inside k8s by using the different Kubernetes plugins~\cite{jenkins-itest}. A comparison of the different features supported by \emph{Bunk8s} with the other tools is shown in Table~\ref{table:comparison}.

Both Istio and Jenkins only support narrow integration testing (\S\ref{sec:integeration_testing}). For Istio, this is possible due to the presence of sidecar containers for pods. On the other hand, both Octopus and \emph{Bunk8s} support broad integration testing which enable the testing of interactions between the application microservices and the backend services. For all the tools shown in Table~\ref{table:comparison}, the developers can use any programming language and test framework for writing integration tests except Istio. Since Istio is an expansion of the Go testing package it only supports writing integration tests in Go. Moreover, Istio is dependent on the usage of Prow~\cite{prow} as the CI/CD system for running the integration tests~\cite{istiotest}. 
Similarly, using Jenkins would require it's introduction into  a project and migration of at least parts of the CI/CD pipeline to Jenkins if not already in use. In contrast, both \emph{Bunk8s} and Octopus can be used with any CI/CD tool that supports Docker containers. A major drawback of using Octopus is that after a test is finished it only reports whether a test failed or was successful and does not extract the test logs from the containers. This prevents an in-depth analysis of test failures~\cite{kyma-octopus}. Moreover, Octopus is not open-source and charges for its services. On the other hand, \emph{Bunk8s} is completely open-source and supports publishing of test results since they are extracted from the test runner pod and are stored on the pipeline runner (\S\ref{sec:sysdesign}). Following this, the supported pipeline mechanisms for test result publishing can be utilized. As a result, the different features supported by \emph{Bunk8s} make it easier for it to be integrated into an existing projects pipeline.

% All tools shown in 

%improve reasoning for broad integeration testing

% Octopus allows developers to define integration tests with custom resource definitions~\cite{crd}. For each test, Octopus deploys a pod to k8s. Furthermore, it supports re-running tests to prevent flaky test results. However, after a test is finished Octopus only reports whether a test failed or was successful and does not extract the test logs from the containers. This prevents an in-depth analysis of test failures~\cite{kyma-octopus}. 

% The Istio testing framework allows developers to test microservice applications that use the Istio service mesh to handle the communication between microservices. The framework is an expansion of the Go testing package and only supports writing integration tests in Go. Moreover, it requires Prow as the CI/CD tool and Github to execute the tests automatically~\cite{istiotest}. With Jenkins the developers can also run integration tests inside of k8s. However, it only supports narrow integration testing, i.e., the test runner containers can only be used to test the interaction of containers within pods. Moreover, using Jenkins would require it's introduction into  a project and migration of at least parts of the CI/CD pipeline to Jenkins if not already in use.

% for using Jenkins it would have to be introduced into the 

\section{Conclusion \& Future Work}
\label{sec:futurework}

% Conclusion
% In the scope of this paper, Bunk8s was developed. It is a testing tool for broad integration testing of microservices deployed in Kubernetes. It is independent of the test framework that is used to write or generate tests, independent of the infrastructure or platform on which Kubernetes runs and supports test result publishing. 
In this paper, we implemented and presented \emph{Bunk8s}, a tool for broad integration testing of microservices in Kubernetes. \emph{Bunk8s} is independent of the test framework used for writing or generating tests, independent of the infrastructure or platform on which Kubernetes runs, and supports test result publishing. \emph{Bunk8s} is well suited for developers planning to write a new test suite as well as for developers who wish to revise and improve the already existing test suite of a microservice application. In the future, we plan to extend \emph{Bunk8s} to support mocking of services. Furthermore, we plan to add support for narrow integration testing in \emph{Bunk8s} to enable testing of multi-container pods. Support for FaaS based CI/CD workloads is another future direction~\cite{jindal2021function}.

\bibliographystyle{IEEEtran}
\thispagestyle{empty}
\bibliography{parallelpgm}

% \listoftodos

\end{document}